\documentclass[twocolumn]{aastex62}

\usepackage{amsfonts}
\usepackage{amsmath}
\usepackage{amssymb}
\usepackage{bm}% bold math
\usepackage{color}
\usepackage{dcolumn}% Align table columns on decimal point
\usepackage{epsfig}
\usepackage{eucal}
\usepackage{float}
\usepackage{graphicx}% Include figure files
\usepackage[latin1]{inputenc}
\usepackage{microtype}
\usepackage{subfigure}
\usepackage{tikz}
\usepackage{ulem}%delete line package
\usepackage{verbatim}
\usepackage{xcolor}

\shortauthors{Jiang et al.}

\begin{document}

\title{Numerical study of inflationary preheating with arbitrary power-law potential and a realization of curvaton mechanism}

\author{Jie Jiang}
\affiliation{Department of Astronomy, School of Physical Sciences, University of Science and Technology of China, Hefei 230026, China}
\affiliation{CAS Key Laboratory for Research in Galaxies and Cosmology, University of Science and Technology of China, Hefei 230026, China}
\affiliation{School of Astronomy and Space Science, University of Science and Technology of China, Hefei 230026, China}

\author{Qiuyue Liang}
\affiliation{Department of Astronomy, School of Physical Sciences, University of Science and Technology of China, Hefei 230026, China}
\affiliation{CAS Key Laboratory for Research in Galaxies and Cosmology, University of Science and Technology of China, Hefei 230026, China}
\affiliation{School of Astronomy and Space Science, University of Science and Technology of China, Hefei 230026, China}
\affiliation{Department of Physics \& Astronomy, University of Pennsylvania, Philadelphia, PA 19104-6396, USA}

\author{Yi-Fu Cai}
\affiliation{Department of Astronomy, School of Physical Sciences, University of Science and Technology of China, Hefei 230026, China}
\affiliation{CAS Key Laboratory for Research in Galaxies and Cosmology, University of Science and Technology of China, Hefei 230026, China}
\affiliation{School of Astronomy and Space Science, University of Science and Technology of China, Hefei 230026, China}

\author{Damien A. Easson}
\affiliation{Department of Physics, Arizona State University, Tempe, AZ 85287-1504, USA}

\author{Yang Zhang}
\affiliation{Department of Astronomy, School of Physical Sciences, University of Science and Technology of China, Hefei 230026, China}
\affiliation{CAS Key Laboratory for Research in Galaxies and Cosmology, University of Science and Technology of China, Hefei 230026, China}
\affiliation{School of Astronomy and Space Science, University of Science and Technology of China, Hefei 230026, China}

%\correspondingauthor{Jie Jiang}
\email{jiejiang@mail.ustc.edu.cn}
%\correspondingauthor{Qiuyue Liang}
\email{qyliang@sas.upenn.edu}
%\correspondingauthor{Yi-Fu Cai}
\email{yifucai@ustc.edu.cn}
%\correspondingauthor{Damien A. Easson}
\email{easson@asu.edu}
%\correspondingauthor{Yang Zhang}
\email{yzh@ustc.edu.cn}

\begin{abstract}
During inflationary preheating, the energy stored in the inflaton field is rapidly converted into excitations of other entropy fields. This stage is characterized by exponential particle production due to parametric resonance and is notoriously difficult to analyze using analytic methods. We develop a detailed numerical simulation to investigate inflationary preheating when the potential of the inflaton is a power-law function with arbitrary power index. To achieve a successful graceful exit from a primordial inflationary phase to a smooth, oscillatory phase during preheating, we assume the inflaton potential reduces to a quadratic function in the infrared regime, which may be regarded as a mass term at low energy scales. With this simplification, our numerical method may be applied to unconventional chaotic inflation models. To demonstrate its validity, we numerically analyze the preheating stage of axion-monodromy inflation which is inspired by string theory. By performing perturbation analyses, our result shows that the power spectrum of primordial curvature perturbation can be dominated by fluctuations of entropy field rather than those of inflaton, which can be regarded as a particular realization of the curvaton mechanism through a preheating process.
\end{abstract}

\keywords{cosmological parameters, inflation, theory}

\section{Introduction} \label{sec:intro}

Reheating is a hypothetical mechanism used to generate a thermal phase of hot big bang expansion from an initially cold and empty primordial Universe. During the reheating phase,  energy is quickly transferred from a primordial matter field, namely, the inflaton scalar field, to other matter fields through interaction terms in the Lagrangian. For inflationary cosmology, this process occurs after the stage of slow-roll inflation, but earlier than the time of thermal equilibrium. The energy transfer from the inflaton to matter field was initially analyzed using first-order perturbation theory and discussed in terms of decaying inflaton quanta into Standard Model particles \citep{Abbott:1982hn, Dolgov:1982th, Albrecht:1982mp}.

It was realized that the coherence of the inflaton condensate at the beginning of the reheating phase can play an important role in the analyses with nonlinear effects \citep{Traschen:1990sw}. Matter generation in the inflaton condensate was studied in a semi-classical analysis involving non-perturbative effects in \citep{Traschen:1990sw}. It was found that a parametric resonance instability plays a crucial role. Parametric resonance after inflation was further studied in \citep{Kofman:1994rk, Shtanov:1994ce} and the particle production process was analyzed in \citep{Kofman:1997yn}. In recent years, this topic has been extensively studied in the literature in the framework of inflationary cosmology \citep{Boyanovsky:1994me, Baacke:1996se, Greene:1997ge, Felder:2000hj, Cormier:2001iw, Shuhmaher:2005mf, Dufaux:2006ee, Abolhasani:2009nb} as well as bounce cosmology \citep{Cai:2011ci, Cai:2013vm, Quintin:2014oea, deHaro:2015wda}. We refer to \citep{Bassett:2005xm, Allahverdi:2010xz, Amin:2014eta} for comprehensive reviews.

While the reheating process can provide a dynamical mechanism for the origin of entropy and particles observed in our universe, details remain somewhat heuristic due to the existence of various uncertainties. For instance, the evolution of the equation-of-state (EoS) parameter strongly depends on the specific construction  \citep{Podolsky:2005bw}. Consequently, one way of probing information about reheating is to track the expansion of the Universe between the moment CMB scales cross the Hubble radius during inflation and the time they re-enter, which connects the energy scale of inflation with the reheating temperature directly. This method was extensively studied in the literature, see for instance, \citep{Liddle:2003as, Martin:2006rs, Martin:2010kz, Adshead:2010mc, Mielczarek:2010ag, Easther:2011yq, Dai:2014jja, Martin:2014nya, Cai:2015soa, Cook:2015vqa, Domcke:2015iaa, Drewes:2015coa, Lozanov:2016hid}.

Nevertheless, such constraints on the reheating temperature strongly rely on measurements of primordial curvature perturbations and relic gravitational waves, from various high precision experiments of cosmic microwave background (CMB) photons and their polarizations \citep{Aghanim:2018eyx, Ade:2015xua}. So far these studies were based on the theoretical predictions of particularly chosen inflation models \citep{Akrami:2018odb, Ade:2015lrj}. Moreover, it was observed that some nonlinear processes, namely parametric resonance effects, which could yield a potential amplification on the amplitudes of primordial density perturbations \citep{Taruya:1997iv, Bassett:1998wg, Bassett:1999mt, Bassett:1999cg, Finelli:2000ya, Tsujikawa:2002nf, Chambers:2007se, Brandenberger:2008if, Bond:2009xx, Bethke:2013aba, Moghaddam:2014ksa, McDonough:2016xvu, Svendsen:2016kvn, BazrafshanMoghaddam:2016tdk, Graef:2017nyv, Gu:2018akj}, have not yet been taken into account in a self consistent way. In particular, if these amplifications have an impact on curvature perturbations due to a conversion from iso-curvature modes, the theoretical predictions for inflation models may be strongly altered or even completely destroyed \citep{Finelli:2000ya, Brandenberger:2008if, Moghaddam:2014ksa, McDonough:2016xvu}. Therefore, it is necessary to numerically examine whether these amplifications occur in specific inflationary models and to further revisit the corresponding predictions for CMB observations.

For those lattice codes of preheating such as LATTICEEASY \citep{Felder:2000hq}, DEFROST \citep{Frolov:2008hy}, PSpectRe \citep{Easther:2010qz} and HLattice \citep{Huang:2011gf}, the cosmological system has been treated as a fixed comoving box. However, in the present study we would like to trace the evolutions of cosmological perturbations at super-Hubble scales so that we could examine whether these modes would continue the evolutions after the Hubble radius crossing. Accordingly, in this article we will use the C++ computer language to develop a new program code to perform the numerical simulation of the inflationary preheating phase, in which the potential of the inflaton is taken to be a power-law function with an arbitrary power index.

To ensure that the oscillatory behavior of the inflaton field around the bottom is continuous and smooth, we introduce a cutoff for the potential in the infrared regime, which behaves as a quadratic potential at low energy scales.The entire background evolution then becomes numerically trackable, even if the underlying inflation model has some unconventional potential forms with singular field derivatives at high energy scales. This treatment is particularly appropriate for a class of the so-called axion-monodromy inflation models inspired by string theory. Here we consider the ordinary chaotic inflation models with quadratic and quartic potentials, and then study particle production in models of power-law potentials with power index being $1$, $2/3$ and $1/2$, respectively.

The present article is organized as follows. In Section~\ref{sec:model}, we introduce the construction of the inflaton potential and introduce the background equations applied in the numerical simulations. In Section~\ref{sec:method}, we give a detailed description of the numerical method we developed to calculate the preheating process. In Section~\ref{sec:application_bg}, we apply the code to several inflationary models to demonstrate the efficiency of the numerical method. In the same Section we firstly consider well-known cases with power index of the potential $p=2$, to demonstrate the validity of the method; and then, investigate some nonconventional models with $p=1$, $2/3$, $1/2$ and $1/3$, which are inspired by models of axion monodromy inflation. Finally, we summarize our results with a discussion in Section~\ref{Sec:conclusions}.

Throughout the article we adopt metric signature $(-, +, +, +)$, use natural units with  $c=\hbar=1$ and use the reduced Planck mass $M_p=1/{\sqrt{8\pi G}}$.

\section{Model construction}
\label{sec:model}

We begin with a homogeneous and isotropic universe, described by the (observationally favored, spatially flat) Friedmann-Lema\^{i}tre-Robertson-Walker (FLRW) metric:
\begin{eqnarray}
 ds^2 = -dt^2 + a^2 d{\rm x}^2 ~,
\end{eqnarray}
where $a$ is the scale factor. Accordingly, the field equations of General Relativity yield the Friedmann equations for the background
\begin{eqnarray}
\label{eq:Friedman}
 H^2 = \frac{1}{3 M_p^2} \rho ~, \quad
 \frac{\ddot{a}}{a} = -\frac{1}{6 M_p^2} (\rho +3p) ~,
\end{eqnarray}
where $H \equiv \dot{a}/a$ is the Hubble parameter characterizing the expansion rate of the Universe, and, the dot represents differentiation with respect to cosmic time $t$. Here $\rho$ and $p$ denote energy density and pressure, respectively.

As matter sources, we take two canonical scalar fields, one responsible for driving a period of slow roll inflation at high energy scale, the other as an entropy field. The Lagrangian is
\begin{equation}
\mathcal{L} = -\frac{1}{2}\partial_{\mu}\phi\partial^{\mu}\phi -\frac{1}{2}\partial_{\mu}\chi\partial^{\mu}\chi - V(\phi, \chi),
\end{equation}
with potential including interaction term,
\begin{eqnarray}
\label{eq:potential}
 V(\phi, \chi) = \lambda M_p^{4-p} \Big[ \big( \phi^2+\mu^2 \big)^{p/2} -\mu^p \Big] + \frac{1}{2} g^2 \phi^2\chi^2 \,.
\end{eqnarray}

\begin{figure}[!h]
\centering
\subfigure{
\begin{minipage}[b]{0.45\textwidth}
\includegraphics[width=1\textwidth]{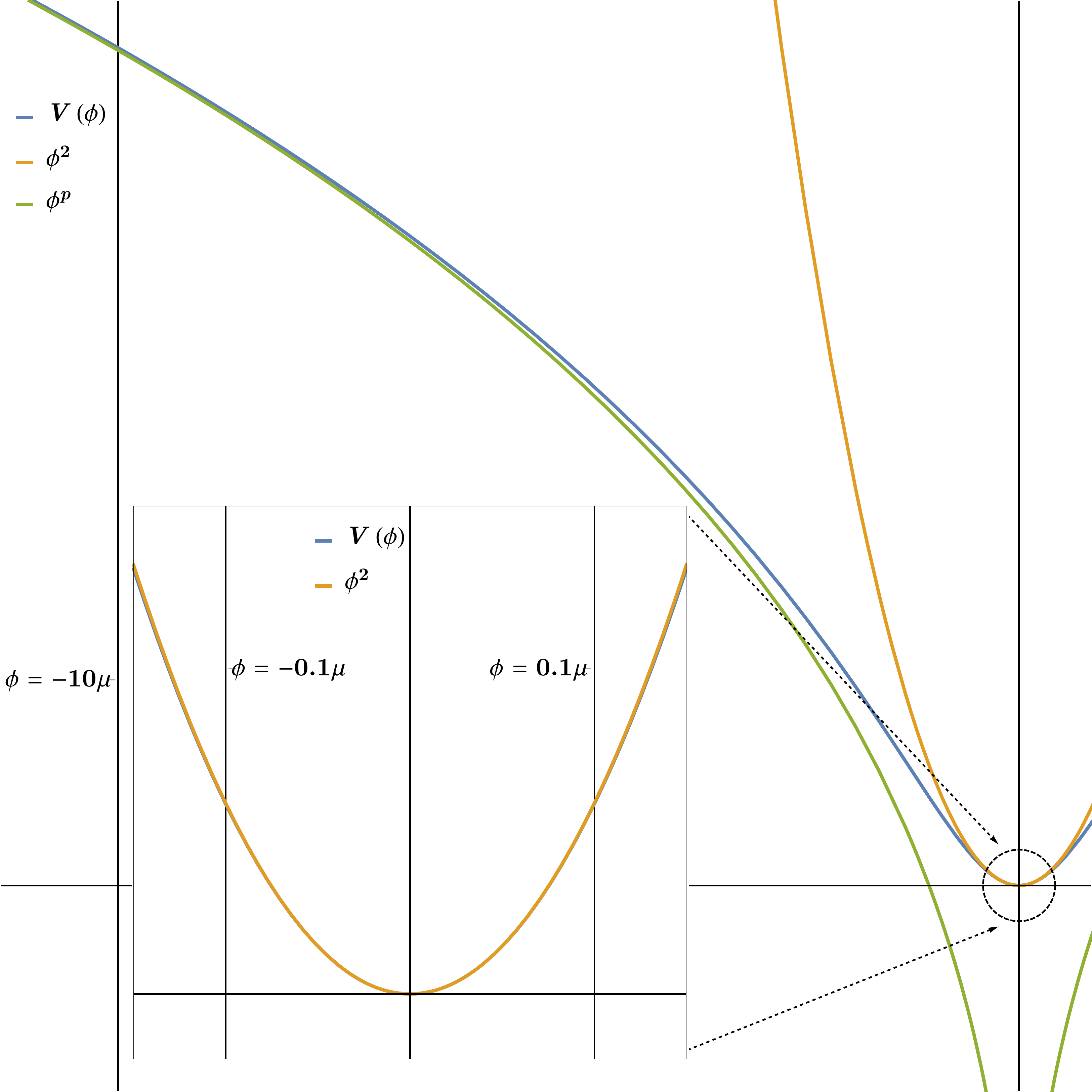}
\end{minipage}
}
\caption{The figure shows the potential as a function of $\phi$. The blue line is the real inflaton potential $V(\phi)$. The green line plots $\phi^p$ and the orange one plots $\phi^2$. As is shown in the figure, when $\phi>10\mu$, $V \propto \phi^p$; while $\phi<0.1\mu$, $V \propto \phi^2$. We adopt $p=1/3$ as an example for illustration.}
\label{fig:potential}
\end{figure}

In the above, $\phi$ denotes the inflaton field and $\chi$ is an entropy field which is assumed to be much lighter than the inflaton, and only excited during the preheating phase. The coefficients $\lambda$ and $g$ are the coupling constants. Here, $p$ is the model-dependent power-law index we discuss in later sections (see also Figure~\ref{fig:potential} for a sketch). In addition, we have introduced a very small mass term $\mu$ in order to ensure the potential behaves smoothly in the infrared regime (in particular, for circumstances when the values of $p$ are smaller than unity).

The background energy density and pressure are,
\begin{eqnarray}
\label{eq:rho&p}
 \rho = \frac{1}{2}\sum_I\dot{\varphi}_I^2+V ~,~~
 P = \frac{1}{2}\sum_I\dot{\varphi}_I^2-V ~,
\end{eqnarray}
where we have ignored spatial gradient terms, $\varphi_I$ represent the inflaton field $\phi$ and the entropy field $\chi$, respectively. The equations of motion for two fields can be derived by varying the Lagrangian with respect to the field variables, respectively:
\begin{eqnarray}
\label{eq:eomphichi}
 \ddot\phi +3H \dot\phi - \frac{\nabla^2}{a^2} \phi + V_{,\phi} =0 ~, \nonumber\\
 \ddot\chi +3H \dot\chi - \frac{\nabla^2}{a^2} \chi + V_{,\chi} =0 ~.
\end{eqnarray}
The field perturbation in spatially flat gauge is the Mukhanov-Sasaki variable
\begin{equation}
 Q_I \equiv \delta\varphi_I+\frac{\bar{\varphi}_I^{\prime}}{\mathcal{H}}\psi ~,
\end{equation}
which is used here for analyzing the perturbation, including the metric perturbation $\psi$, where $\mathcal{H}$ is the comoving Hubble parameter. The equations of motion of the perturbation fields are
\begin{align}
\label{eq:eomQphichi}
 \ddot{Q}_I &+3H\dot{Q}_I -\frac{\nabla^2}{a^2}Q_I + ~\nonumber \\
 &\sum_J \bigg[ V_{,IJ} -\frac{1}{M_p^2a^3} \frac{d}{dt} \Big( \frac{a^3}{H} \dot{\varphi}_I \dot{\varphi}_J \Big) \bigg] Q_J = 0 ~,
\end{align}
where the comma denotes the derivative with respect to the field $\phi$ or $\chi$. After operating the Fourier transform to Equation \eqref{eq:eomQphichi}, the partial derivative with respect to space coordinate, $-\nabla^2$, changes to the comoving wavenumber $k^2$ in Fourier space. The Partial Differential Equations (PDE) then become Ordinary Differential Equations (ODE), simplifying computations.

During inflation, the contribution of the $\chi$ field is negligible since the $\chi$ energy density is completely diluted by the exponential expansion of the universe without fine-tuning parameters. Consequently, it is natural to take $\chi_I = \dot{\chi}_I = 0$ as the initial conditions at the beginning of inflation. Following the traditional slow-roll paradigm, we introduce the slow-roll parameter
\begin{equation}
 \epsilon = \frac{M_p^2}{2} \Big[\frac{V(\phi,0)^{\prime}}{V(\phi,0)}\Big]^2 ~,
\end{equation}
and inflation ends when $\epsilon=1$. We set the initial conditions of the perturbations to be the Bunch-Davies vacuum. As the power index $p$ of the inflaton potential changes, the initial conditions of $\phi$ and $\dot{\phi}$ must change in order to achieve a sufficient number of e-foldings, $N \approx 50\sim 60$. Furthermore, in order to ensure the potential behaves like $\phi^p$
\begin{equation}
 V \approx\lambda M_{p}^{4-p}[\phi^p-\mu^p] ~,
\end{equation}
during the inflation, we require that $|\phi| \gg 10\mu$, restricting
\begin{equation}
 \phi_{\epsilon=1} \geq 10\mu ~.
\end{equation}
Thus, for different $p$, we can yield different upper limits for $\mu$. Note that, during inflation, the effective mass of the inflaton is given by
\begin{equation}
 m_{\phi_{\text{eff}}}^2 = 2 \lambda M_{p}^{4-p} \phi^{p-2} ~.
\end{equation}

After inflation, the universe is dynamically driven by the oscillations of $\phi$ about the vacuum, which is called preheating. The energy transfers from the inflaton to the matter field, exciting relativistic particles. In order to transfer the energy, the effective mass of the inflaton must on average continually decrease. However, for those string theory inspired axion-monodromy models where $p<2$, the inflaton becomes heavier and heavier, yielding an inefficient preheating process. In our model, when $|\phi|$ is smaller than $0.1\mu$, the potential $V$ becomes
\begin{equation}
 V(\phi,\chi)=\lambda\frac{p}{2}\frac{\mu^{p-2}\phi^2}{M_{p}^{p-4}}+\frac{1}{2}g^2\chi^2\phi^2+O(\phi^4) ~.
\end{equation}
The effective mass of inflaton is
\begin{equation}
 m_{\phi_{\text{eff}}}^2=p\lambda M_{p}^{4-p}\mu^{p-2} ~,
\end{equation}
is constant instead of becoming heavier. The graceful exit from $\phi^p$ to $\phi^2$ results in the constant effective mass of the $\phi$ field, and efficient particle production during the preheating phase. During this time, the EoS parameter is as during matter-domination, so that $a \propto t^{2/3}$, $H = \frac{\dot{a}}{a} = \frac{2}{3t}$, and the equation of motion of the inflaton becomes
\begin{equation}
 \ddot{\phi} + \frac{2}{t}\dot{\phi} + m_{\phi_{\text{eff}}}^2\phi=0 ~.
\end{equation}
This equation of motion is solved by
\begin{equation}
 \phi = \frac{\Phi_0}{m_{\phi_{\text{eff}}}t} \sin(m_{\phi_{\text{eff}}}t) ~,
\end{equation}
where $\Phi_0$ is the coefficient when the inflaton enters the $V \approx \phi^2$ regime. Since the coupling is $1/2g^2\phi^2\chi^2$ in the model, we have a kind of scattering process, so the matter field effective mass $m_{\chi_{\text{eff}}}^2$ should be lighter than the effective mass of the inflaton $m_{\phi_{\text{eff}}}^2$. Since during the $\phi^2$ preheating period, $\phi<0.1\mu$, so $\Phi<0.1\mu$, and
\begin{equation}
 m_{\phi_{\text{eff}}}^2 = p\lambda M_{p}^{4-p}\mu^{p-2} > m_{\chi_{\text{eff}}}^2 = g^2\Phi^2 ~,
\end{equation}
we choose the upper limit of $\Phi$, so that
\begin{equation}
 g<\sqrt{100 p \lambda M_{p}^{4-p} \mu^{p-4}} ~.
\end{equation}
The coupling parameter $g$ should be carefully chosen for parametrically resonance. Since the potential is proportional to $\phi^2$ during preheating, the condition for parametric resonance is quite similar to that of pioneer work on $\phi^2$ inflation \citep{Kofman:1997yn}. We rescale the matter field $\chi_k$ with $X_k = a^{3/2}\chi_k$, and rewrite the equation of motion:
\begin{equation}
 \ddot{X}_k+\omega_k^2X_k=0 ~,
\end{equation}
where $\omega^2_k = \frac{k^2}{a^2} + g^2\phi^2 - \frac{9}{4}H^2 - \frac{3}{2}\dot{H}$, the last two terms vanish as the EoS parameter approaches zero, and
\begin{equation}
 \omega^2 = \frac{k^2}{a^2} + \frac{g^2\Phi_0^2}{m_{\phi_{\text{eff}}}^2 t^2} - \frac{g^2\Phi_0^2}{m_{\phi_{\text{eff}}}^2 t^2}\cos(2m_{\phi_{\text{eff}}} t) ~.
\end{equation}
In the standard Mathieu function form \citep{McLachlan:1947} we have,
\begin{equation}
 X_k'' + [A_k-2q\cos(2 z)] X_k=0 ~,
\end{equation}
where $z = m_{\phi_{\text{eff}}} t$, a prime denotes the derivate with respect to $z$, and $A_k = \frac{k^2}{m_{\phi_{\text{eff}}}^2 a^2} + \frac{g^2 \Phi_0^2}{m_{\phi_{\text{eff}}}^4 t^2}$, $q = \frac{g^2\Phi_0^2}{4 m_{\phi_{\text{eff}}}^4 t^2}$.
We use $q = \frac{g^2\Phi^2}{4 m_{\phi_\text{eff}}^2}>1 $ to determine the broad resonance, where $\Phi$ is the initial amplitude. We see that when $\mu$ decreases, the effective resonance area decreases, thus the lower limit for $g$, $g > 2 \Phi m_{\phi_\text{eff}}$, increases, indicating stronger coupling.

In order to numerically investigate the evolution of the matter field $\chi$ in this phase, we introduce a small, homogenous and isotropic perturbation $\delta \chi$ as the initial value for $\chi$ and $d\chi$ at the beginning of preheating. The background homogeneous part of the $\chi$ field, though very small, can be interpreted as the condensate matter field.  The inhomogeneous part, $\chi_k$, can, in turn, be interpreted as the matter particles with different comoving wavenumbers excited by the scattering with the inflaton field. It is important to distinguish $\chi_k$, and the spatially flat field perturbation, the Mukhanov-Sasaki variable $Q_\chi$, which includes the metric perturbation in our code. The occupying number $n_k$ of the $\chi$ field is defined by $X_k = a^{3/2}\chi_k$ instead of $Q_\chi$,
\begin{equation}
 n_k = \frac{\omega_k}{2} \left(\frac{|\dot{X_k}|^2}{\omega_k^2} +|X_k|^2 \right) -\frac{1}{2} ~,
\end{equation}
and the comoving curvature perturbation
\begin{equation}
 \mathcal{R} = H \frac{Q_\phi \dot{\phi} + Q_\chi \dot{\chi}}{\dot{\phi}^2 + \dot{\chi}^2} ~,
\end{equation}
where the curvature perturbation mainly depends on $Q_\phi$ and $Q_\chi$. Previously, people didn't include the metric perturbation when they try to consider the change of curvature perturbation caused by matter field.  Instead, they introduce the back reaction afterward. However, due to the improving ability to calculate, we are able to consider the metric perturbation at the beginning, which highly improves the precision.

\section{Description of the numerical method}
\label{sec:method}

In this section, we present a detailed description of the numerical method we apply to the preheating analysis. Previous codes only considered the field fluctuations instead of including metric perturbations, which leads to an ambiguous definition between the curvature perturbation and the occupation number. In our code, we separate the field perturbation and the Mukhanov-Sasaki variable, making the definition clear. In addition, the complicated non-linear process in lattice simulations is unnecessary, since the linear perturbation theory of two fields would, in general, modify the power spectrum of the curvature perturbation. We write an MPI-version to run the evolution for different wave vector in parallelization and get the power-spectrum all at once. We write the code for inflationary preheating in the C++ computer language and hope to revisit the inflationary predictions of the $(n_s-r)$ contour plot. The numerical algorithm adopted in our code is based on the 4th order Runge-Kutta approach, which is comparably precise and fast.

We introduce the application of 4-th order Runge-Kutta method in Section~\ref{sub:4order} and discuss the structure of the code together with the initial values in Section~\ref{sub:structure}.

\subsection{4-th Order Runge-Kutta Method}
\label{sub:4order}

Here we give a brief introduction to 4-th order Runge-Kutta method for our ODE set, (see Equations~\eqref{eq:Friedman}, \eqref{eq:eomphichi} and \eqref{eq:eomQphichi}). For a differential equation $\frac{dy}{dx}=f(x,y)$, the next step of $y_{n+1}$ is
\begin{align*}
 &y_{n+1}=y_n+\frac{h}{6}[K_1+2K_2+2K_3+K_4]\\
 &K_1=f(x_n,y_n)\\
 &K_2=f(x_n+\frac{1}{2}h,y_n+\frac{1}{2}hK_1)\\
 &K_3=f(x_n+\frac{1}{2}h,y_n+\frac{1}{2}hK_2)\\
 &K_4=f(x_n+h,y_n+hK_3)
\end{align*}
where $h$ is the time step. Using this method the numerical computation can converge very fast and remain high precision.

\subsection{The structure of the code}
\label{sub:structure}

We first calculate the inflationary part. The matter field $\chi$ has almost diluted away so it is appropriate to set $\chi = \dot{\chi} \equiv 0$ during inflation. $\phi_I$ and $\dot{\phi}_I$ is determined by $N_{inf}$, the e-folding number, to ensure a theoretically viable phase of inflation. The Mukhanov-Sasaki variable $Q_\phi$ and $Q_\chi$ are introduced at the beginning of the inflation, where the Bunch-Davies vacuum gives the initial values for these two variables. We refer to Appendix~\ref{app:Q} for the detailed description.

With the strong assumption that cosmological perturbations are not altered throughout the whole preheating phase, one can directly restrict the value of the model parameter $\lambda$ by applying observational data from CMB experiments. Namely, one can (roughly) set the e-folding number and the amplitude of power spectrum of primordial curvature perturbations to be,
\begin{eqnarray}
 N_\text{inf} &\equiv& \ln{a_\text{end}} - \ln{a_\text{I}} \in [50, 60] ~, \nonumber\\
 P_R &=& \frac{H^2}{8\pi ^2 \epsilon M_p^2} \approx 2 \times 10^{-9}~,
\end{eqnarray}
where $\epsilon$ is the slow-roll parameter. When $\epsilon \sim \mathcal{O}(1)$, inflation ends, and $\phi_\text{end}$ is used for determining $\mu$ by setting the upper limit $\mu_\text{max} = 0.1 |\phi_\text{end}|$.

\begin{table}[h]
\label{ini}
\centering
\begin{tabular}{|c|c|c|c|c|c|c|c|}
\toprule
$p$   &   $~\phi_\text{I} ~ (M_p)$ & $\lambda$  &$\mu_\text{max}$  & $g$ & $q$\\
\hline
~$1/3$~ &  -7.9   & ~$3.7 \times 10^{-10}$~ &  ~0.043~ &~0.0868  &~1.48968~\\
\hline
$1/2$ &   -9.2  &~ $3.7 \times 10^{-10}$~ &~0.05~ & 0.075864 & 2.17388\\
\hline
$2/3$ & -10.2  &  ~$3.2 \times 10^{-10}$~  &~0.058 ~ & 0.057 & 2.8855 \\
\hline
$1.0$ & -12.1   & ~$2.0 \times 10^{-10}$~   &~0.078~  & 0.0288 & 4.92016 \\
\toprule
$2.0$ &  -17.4   &  ~ $1.7 \times 10^{-11}$~ & 0& $0.003291$ & $\setminus$\\
\hline
%$4.0$ &   -24  &  ~$3.5 \times 10^{-12}$~ & 0 & $4\times 10^{-7}$ & $\setminus$\\
%\hline
\end{tabular}
\caption{Model parameters and initial conditions during inflation.}
\end{table}

After inflation, we take the ending values of
\begin{equation}\nonumber
(\phi~,\dot{\phi}~, Q_\phi~, \dot{Q}_\phi~, Q_\chi~, \dot{Q}_\chi)_{|\text{end}}
\end{equation}
as the initial conditions of the preheating phase $(\phi~, \dot{\phi}~, Q_\phi~, \dot{Q}_\phi~, Q_\chi~, \dot{Q}_\chi)_{|\text{p}}$. For the $\chi$ field, we introduce the Bunch-Davies Vacuum as the initial value $\chi_{\text{p}}(k)$ during the preheating period (See Appendix~\ref{app:chi}). The reason we do not let $\chi$ evolve through the entire process is due to its complete dilution due to the hierarchy between $\chi$ and $\phi$ during inflation. We determine the coupling constant $g$ by the semi-analytical method.

During preheating, the inflaton $\phi$ oscillates near $\phi =0 $ with the potential $V \sim \phi^2$, thus we can approximate $\phi \sim \Phi \sin{(m_{\phi_\text{eff}} t)}$, where $\Phi$ is the amplitude and $m_{\phi_\text{eff}} = \sqrt{ p \lambda M_p^{4-p}\mu^{p-2}}$ is the effective mass during preheating. We take $\Phi \approx 0.1 \mu$ for simplicity, where $0.1 \mu$ represents the resonance area. The equation of motion of $X_k(t) = a^{3/2}(t) \chi_k(t)$ is,
\begin{eqnarray}
\label{eq:Xk}
 \ddot{X_k} &+& \Big( \frac{k^2}{a^2} + g^2\Phi^2(t) \sin^2{m_{\phi_\text{eff}} t} \Big) X_k =0 ~,
\end{eqnarray}
where we have neglected the $(-\frac{9}{4}H^2 - \frac{3}{2}\dot{H})$ term since the background is matter-dominated. One can straightforwardly rewrite Equation~\eqref{eq:Xk} in the form of the Mathieu equation as follows,
\begin{align}
\label{eq:Xkz}
 & X_k'' + (A_k - 2q \cos{2z}) = 0 ~, \nonumber\\
 & A_k = \frac{k^2/a^2}{m_{\phi_\text{eff}}^2} +2q ~,~~ q = \frac{g^2\Phi^2}{4m_{\phi_\text{eff}}^2}\approx \frac{g^2 \mu^{4-p}}{400 p \lambda M_p^{4-p} } ~,
\end{align}
where the prime denotes the derivative with respect to $z=m_{\phi_\text{eff}} t$. The parameter $q$ is often used to distinguish the narrow resonance ($q\ll 1$) and the broad one ($q\gg 1$). The narrow resonance is inefficient and only suitable for narrowband near $k \approx m_{\phi_\text{eff}}$. In the present article, we analyze $q \sim 1.5$ which induces broad resonance for all long wavelength modes. These parameters are capable of efficiently producing matter particles, changing $P_\mathcal{R}$, the power spectrum of curvature perturbation,  by adding entropy perturbation and also restricting $P_\mathcal{R}$ within the observational limit.

The moment when the energy densities of the two fields are equivalent, i.e., $\rho_\phi \simeq \rho_\chi$ denotes the end of preheating. By comparing $\frac{Q_{\chi|\text{end}}}{Q_{\chi|\text{p}}}$, one can determine whether efficient preheating occurs. Practically speaking, we run the code until the order of $Q_\chi$ stops rising and calculate the curvature perturbation there.

Since we focus on the linear perturbation, we use MPI to run parallel computation for different comoving wavenumbers. Choosing $10^{-4} ~ \text{Mpc}^{-1}< k < 10^{-1} ~ \text{Mpc}^{-1}$ and converting them into Planck units, we get the range for comoving wavenumber as follows,
\begin{eqnarray}
\label{eq, krange}
  k \in [ 1\times10^{-60},1\times10^{-57}]~ l_p^{-1}.
\end{eqnarray}
However, the observational range of $k$ is based on the fact that our current scale factor $a_0 =1$. In our code, we set $a_\text{I} = 1$ at the beginning of inflation for calculation convenience. To make $k/a$ in the same order with the observation, we need to adjust the range of $k$ in our code. Supposing the preheating temperature $T_{\text{re}} \simeq 10^{13}~ \text{GeV}$ and from the relation $\frac{T_{\text{re}}}{T_{0}} = \big(\frac{43}{11 g_{s,\text{re}}} \big)^{1/3} \frac{a_0}{a_{\text{re}}}$ \citep{Dai:2014jja},
we can estimate the ratio of $a_0$ to $a_{\text{re}}$ by $10^{13}~ \text{GeV}/10^{-4}~ \text{eV} \sim 10^{26}$.  The initial condition should be chosen to ensure $k / aH \gg 1$ at the beginning of inflation so that every wave vector is inside the Hubble radius. That is,
\begin{equation}
	\frac{k_{\text{min}}}{e^{- N_{k}} e^{- N_{\text{re}}}} \gg H_{\text{ini}},
\end{equation}
where $N_{k}$ is the e-folding number during inflation, $N_{\text{re}}$ is the e-folding number from the start of preheating to today and $H_{\text{ini}}$ is the initial Hubble parameter. Since we have already have $N_{\text{re}} = \ln(10^{26})$, we can use the above conditions to get the right range of the wavenumbers for different case of $p$.

Generally speaking, the scale factor will bump $e^{70} \sim 10^{30}$ during inflation no matter which $p$ we choose. Together with the additional $26$ orders after the inflation, the total e-folding number is about $55$. It is proper to choose the wavenumber range for our code to be $(10^{-5}~l_p^{-1}, 10^{-2}~l_p^{-1})$. In the following sections, we show the pivot scale, $k=10^{-3}~l_p^{-1}$, for illustration.

To test precision, we use two different scale factor evolution methods. One is $\dot{a}/a = H = \sqrt{\rho/(3M_p^2)}$, and the other $\ddot{a}/a = -(\rho + 3p)/(6M_p^2)$. They give equivalent results for the following sections.

\section{Applications to specific examples of the background field evolution}
\label{sec:application_bg}

We now apply the numerical method introduced above to our potentials with different polynomial power laws. We begin with the classic preheating model (eg, $p=2$), which has been analyzed thoroughly~\citep{Kofman:1997yn}. From the plot, we will show that our method agrees with previous results. We then consider other cases (e.g, $p = 1,~ 2/3,~ 1/2,~ 1/3$ ) to analyze the process of preheating for small field inflation.

\subsection{Regular chaotic inflation with $p=2$}

Following \citep{Bassett:2005xm}, we plot the production of particles in for $p = 2$. When choosing the proper initial conditions, the preheating period is efficient, producing enough matter particle. Our result matches the work of \citep{Kofman:1997yn} while they chose the mode $k=0$, which can be understood as producing the condensate $\chi$ particles.

\begin{figure}[!h]
\centering
\includegraphics[width=0.45\textwidth]{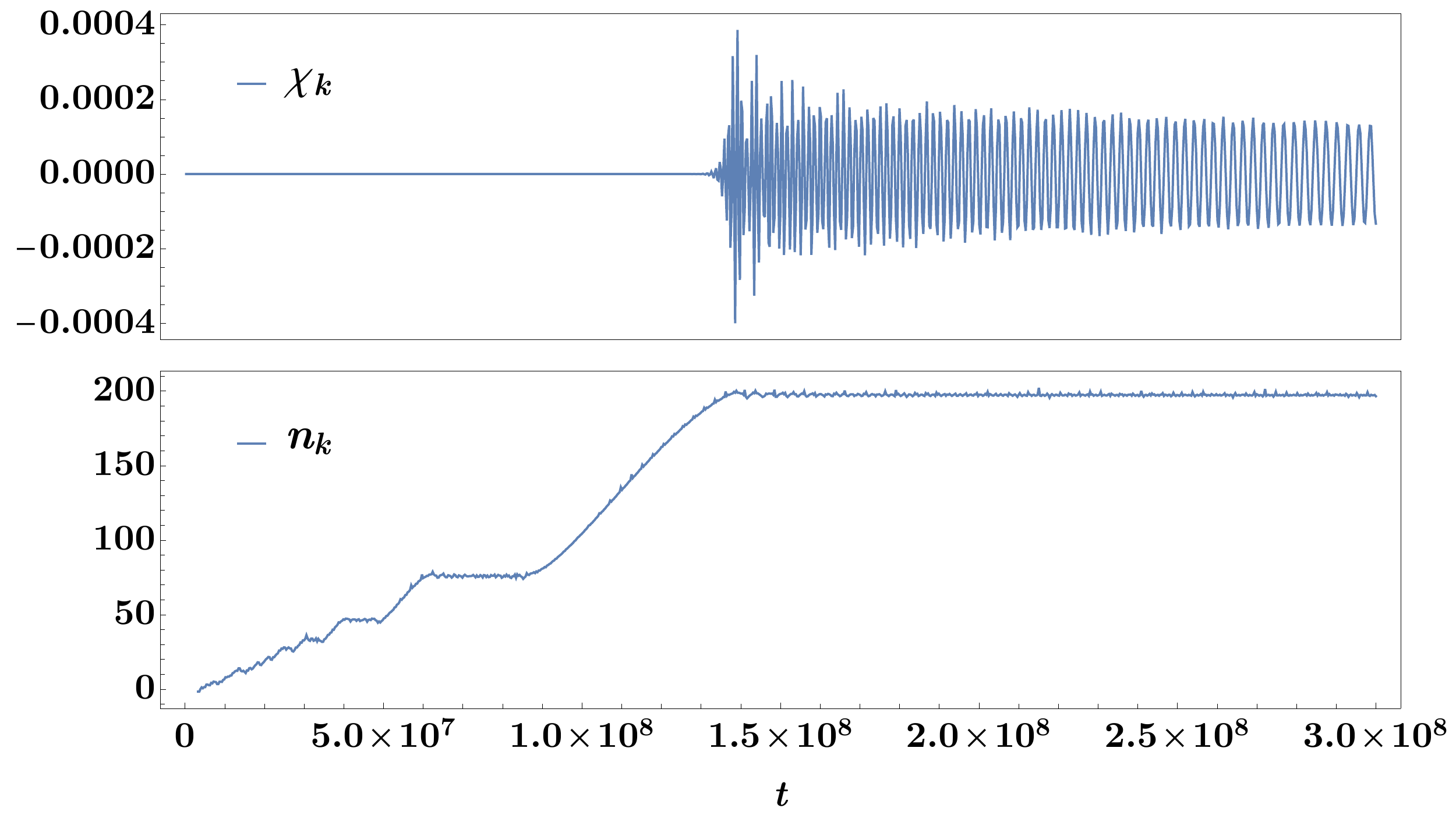}
\caption{Evolution of $\chi$ field(upper) and the comoving number density(lower) with parameters $p=2$, $\lambda = 1.7\times 10^{-11}$, $g = 0.003291$, $k = 10^{-3}$. The comoving wavenumber $k$ is in Planck units.}
\label{fig:p=2-2}
\end{figure}

From Figure~\ref{fig:p=2-2}, we see that for the case of $p=2$, the $\chi$ field can experience the resonance excitation in a chosen parameter band, and the physical particle number increases when $\chi$ is resonating. However, potentials with $\phi^2$ or $\phi^4$ are now significantly disfavored by the CMB data measured by the Planck satellite \citep{Akrami:2018odb}, motivating consideration of potentials with $p<1$. The key point is to see whether there will be enough particle production and whether the curvature perturbation predictions are significantly changed.

\subsection{Nonconventional models with $p=1, 2/3, 1/2, 1/3$}

\begin{figure}[!h]
\centering
\subfigure{
\begin{minipage}[b]{0.48\textwidth}
\includegraphics[width=\textwidth, keepaspectratio=true]{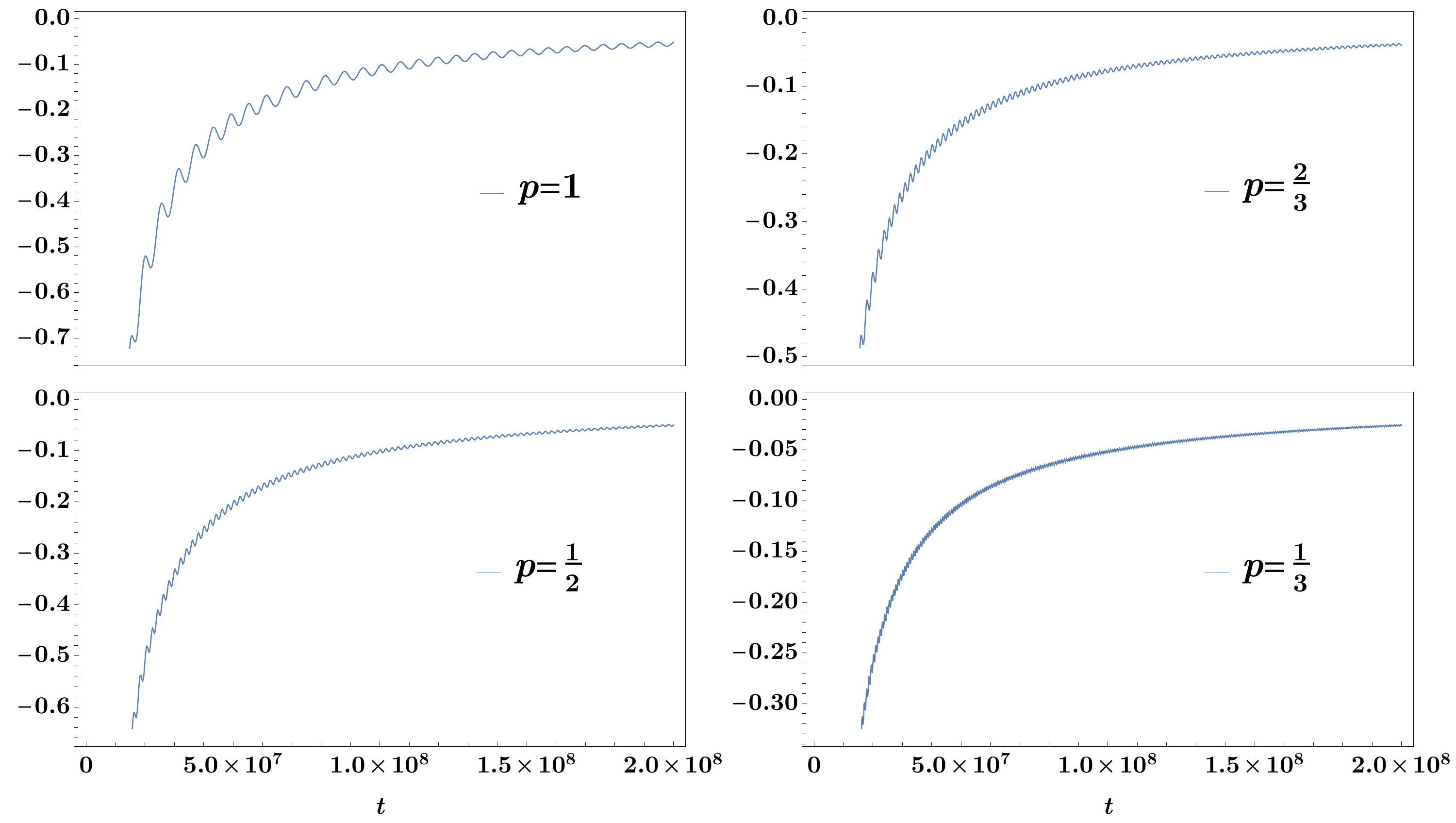}
\end{minipage}
}
\caption{Evolutions of the effective EoS parameter $w \equiv P/\rho = (\frac{1}{2}\sum\dot{\varphi}_I^2 -V )/(\frac{1}{2}\sum\dot{\varphi}_I^2 + V )$ for $p=1$, $2/3$, $1/2$ and $1/3$, which are shown in the longitudinal axes, as functions of the cosmic time $t$. All curves converge to $0$ finally, which correspond to the matter-dominated preheating phase. The cosmic time $t$ is in Planck time units.}
\label{fig:w}
\end{figure}

Now we apply the code to analyze the preheating process when $p<1$. As has been discussed previously, we ensure a smooth transition from $\phi^p$ to $\phi^2$ by fixing the upper limit of $\mu$. As a result, we can make the approximation $\phi \sim \sin{m_{\phi_\text{eff}}t}$, which simplifies the analytical calculation of $\chi$ fields. It can be seen in the evolutions of EoS, which are shown in Figure~\ref{fig:w}. Different potentials ultimately reduce to the case of ($p=2$, $w=0$) at the end.

It is well known that for $\phi^2$, the particle production is sufficient. As our potentials always come back to $\phi^2$, the particle production should also be sufficient. Indeed, our numerical results indicate the $p<1$ cases efficiently produce particles.

\begin{figure}[!h]
\centering
\subfigure{
\begin{minipage}[b]{0.45\textwidth}
\includegraphics[width=\textwidth,keepaspectratio=true]{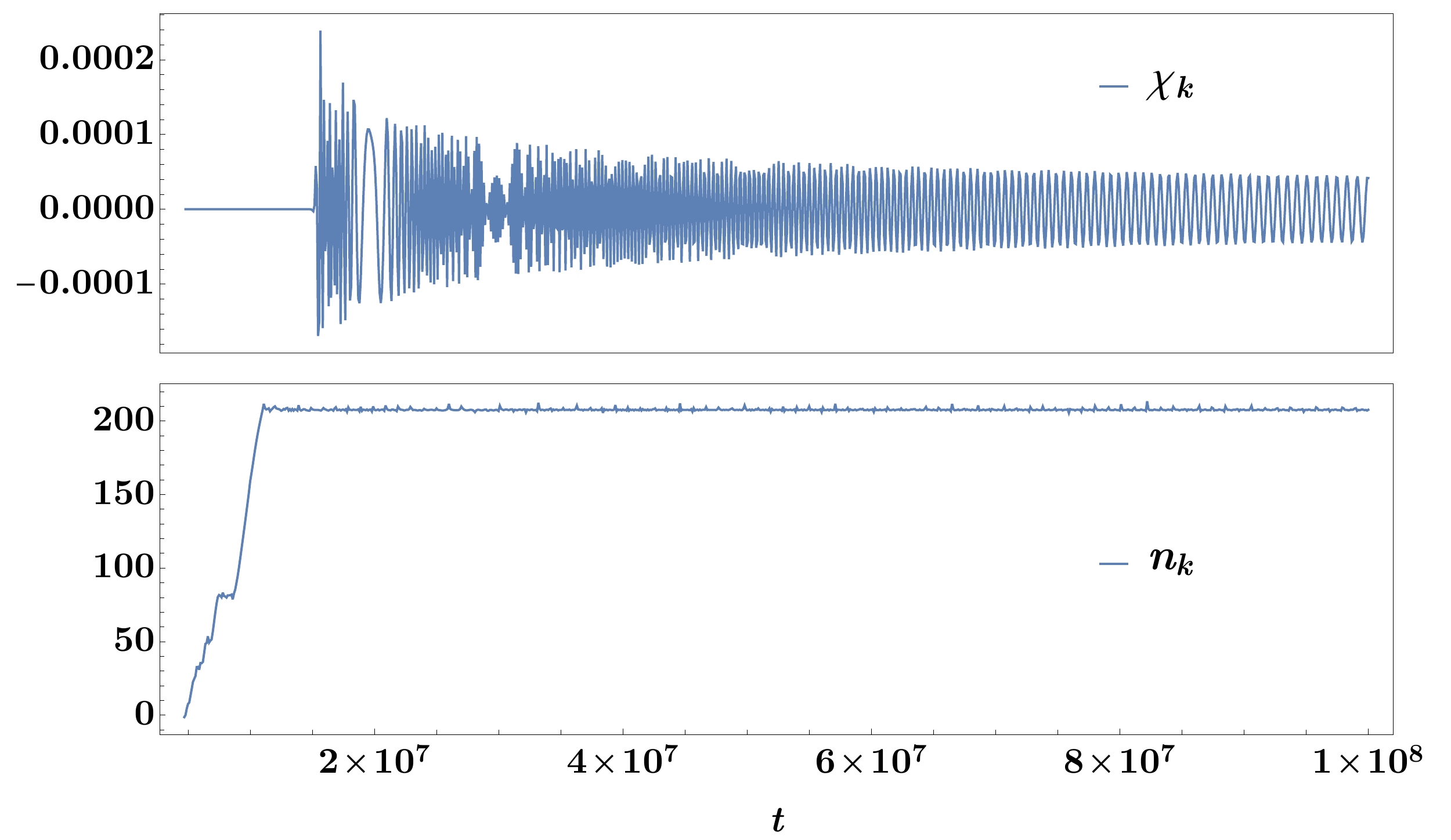}
\end{minipage}
}
\caption{Evolutions of $\chi$ field (upper) and the comoving number density (lower) with parameters $p=1/2$, $\lambda = 3.7\times 10^{-10}$, $g = 0.075864$, $\mu = 0.05$, $k = 10^{-3} ~l_p^{-1}$. The cosmic time $t$ is in Planck time units.}
\label{fig:0.5-2}
\end{figure}

Figure~\ref{fig:0.5-2} shows the $\chi$ field and occupying number in the case of $p = 1/2$. We can see that at the early time, the comoving particle number density increases while $\chi$ is resonating, and then remains unchanged in the comoving universe. The analysis recovers the results found in the $\phi^2$ case, as expected.

\section{Applications to specific examples of the perturbation and power spectrum}
\label{sec:application_pert}

One of the most powerful observational results of the CMB data is the $(n_s -r)$ plot which almost rules out $\phi^2$ and $\phi^4$ models. However, it is unfair to avoid considering the significant preheating process when approximating the spectral index $n_s$. The effect that the $\chi$ field acting as an entropy field to unfreeze the curvature perturbation on super-Hubble scales cannot be neglected. In this section, we analyze how the second field influences the curvature perturbation and how it may correct the predictions on the power spectrum of single field inflation.

\subsection{Regular chaotic inflation with $p=2$}

We begin with an analysis of the standard quadratic model, $V\sim \phi^2$.
\begin{figure}[!h]
\centering
\includegraphics[width=0.45\textwidth,keepaspectratio=true]{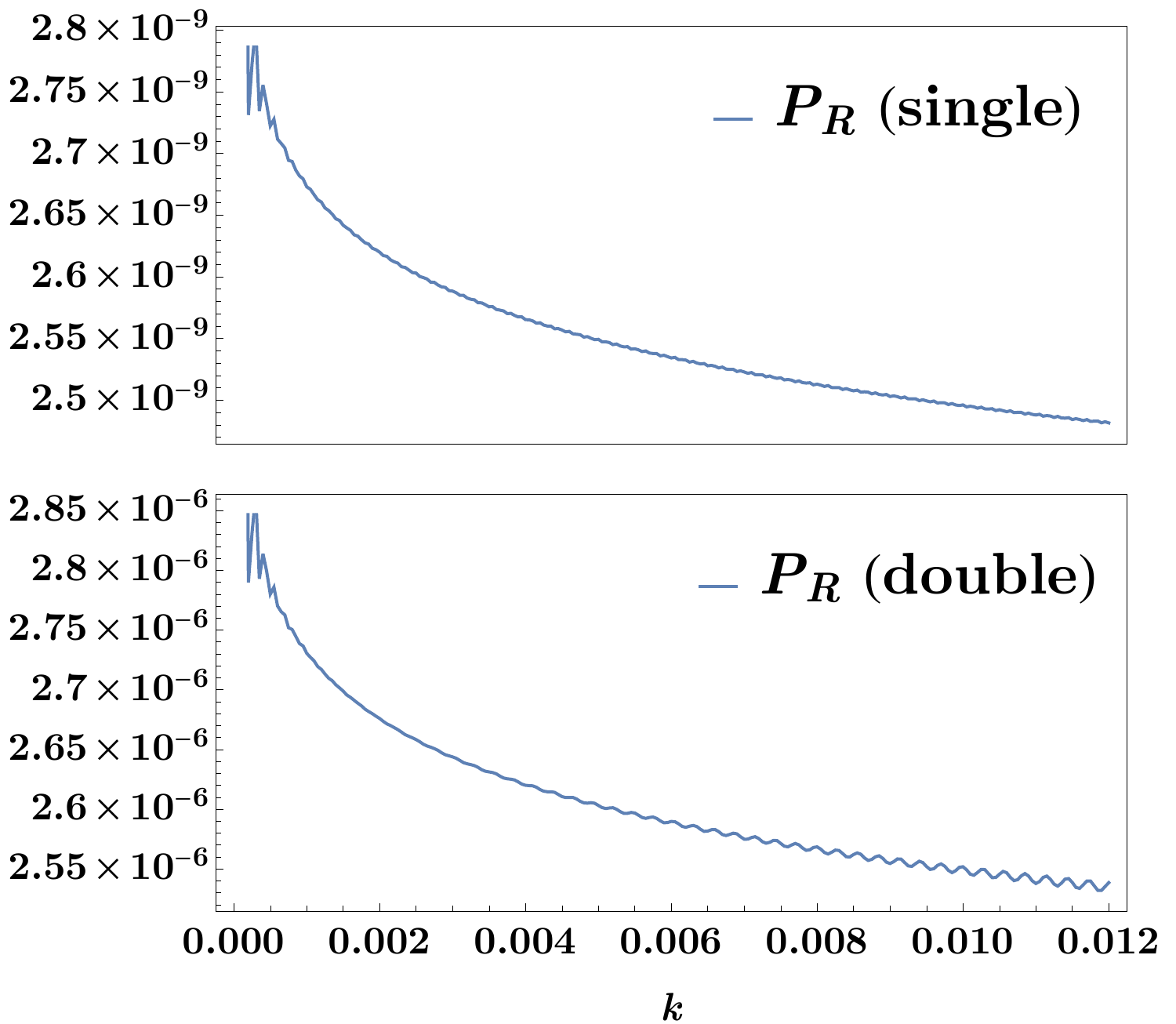}
\caption{We show the power spectrum of the double field curvature perturbation(lower) has an overall amplification compared to the single field one(upper) with parameters $p=2$, $\lambda = 1.7\times 10^{-11}$, $g = 0.003291$. The comoving wavenumber $k$ is in Planck units.}
\label{fig:power-2}
\end{figure}

Figure~\ref{fig:power-2} shows that the power spectrum of the double field curvature perturbation is also nearly scale-invariant, like that of the single field curvature perturbation; however, due to particle production, the magnitude of the curvature perturbation rises by nearly 3 orders of magnitude. In other parameter bands, this amplification will be even more significant and uncontrolled. This provides a new way to constrain the parameter space. Moreover, the analysis demonstrates that the curvature perturbation is not frozen after exiting the Hubble horizon and will keep evolving driven by entropy perturbations. It is therefore inappropriate to use the {\it single field curvature perturbation} when extra matter fields are introduced to make efficient particle production during preheating. One ought to consider a combination of both fields when analyzing the CMB information from inflationary predictions, which will be addressed in the follow-up project.

It is interesting to note that, for a sizeable parameter space as illustrated in the above numerical estimate, the power spectrum of primordial curvature perturbation remains nearly scale-invariant, but the amplitude can be altered significantly due to the contribution of the fluctuations of the entropy field. This scenario coincides with the so-called curvaton mechanism \citep{Mollerach:1989hu, Linde:1996gt, Lyth:2001nq, Moroi:2001ct, Enqvist:2001zp}, which requires at least two primordial fields with one being the inflaton and the other the curvaton field. When the curvaton is subdominant it only provides entropy perturbations during inflation \citep{Lyth:2002my}, and afterward, these entropy perturbations can be converted to curvature perturbations through various dynamical processes \citep{Sasaki:2006kq, Easson:2007dh, Li:2008fma, Huang:2008bg, Huang:2008zj, Cai:2008if, Cai:2009hw, Kobayashi:2009cm, Zhang:2009gw, Gong:2009dh, Cai:2010rt, Cai:2011zx, Alexander:2014uaa, Addazi:2016rnz} so that the curvaton field can start to dominate the universe. For traditional curvaton models, the universe has to enter the standard thermal history after the curvaton decays. However, in our scenario, the curvaton mechanism itself is achieved by the preheating process.

\subsection{Nonconventional models with $p=1, 2/3, 1/2, 1/3$}

\begin{figure}[!h]
\centering
\subfigure{
\begin{minipage}[b]{0.45\textwidth}
\includegraphics[width=\textwidth]{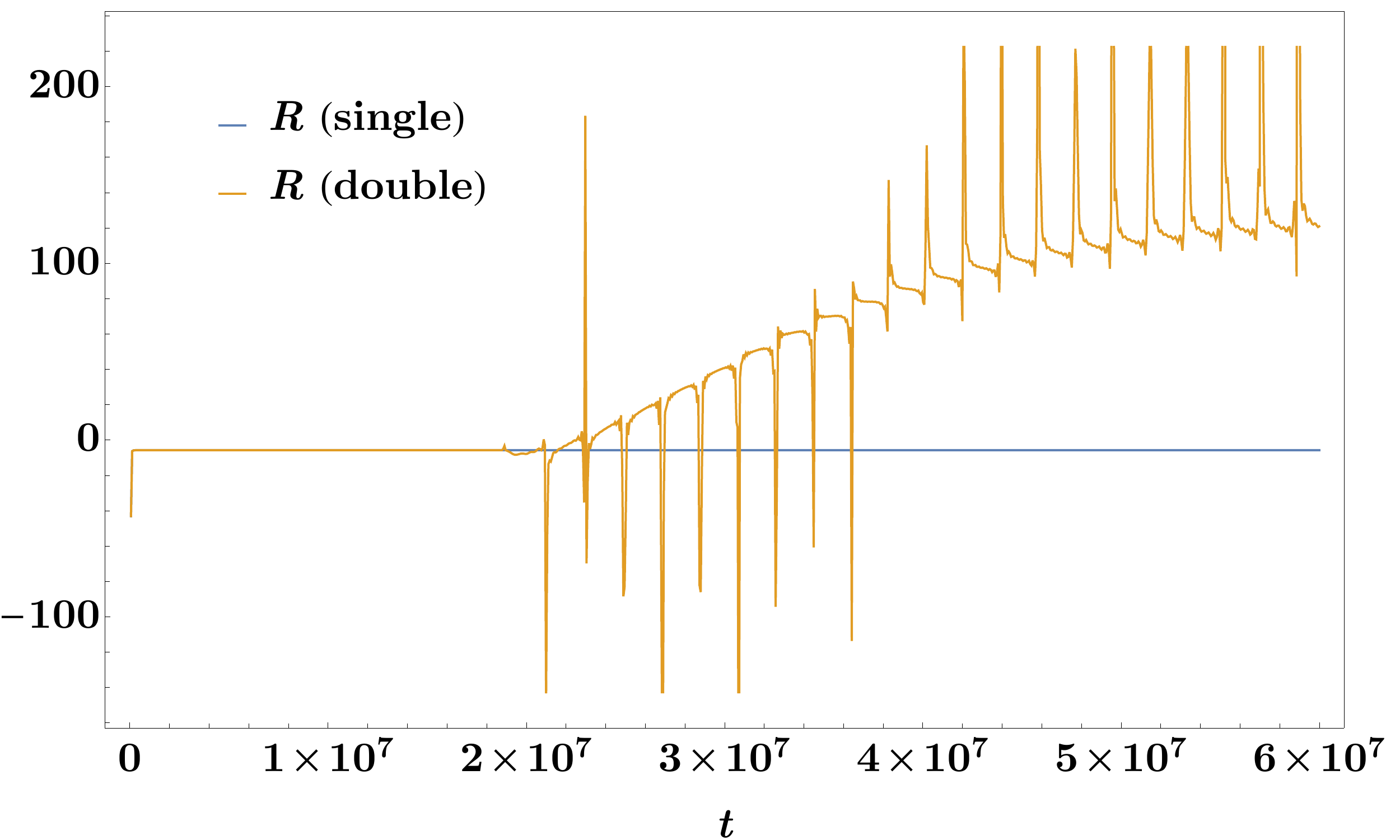}
\end{minipage}
}
\caption{Evolutions of the curvature perturbation of single field case(blue) and double field case(orange) with parameters $p=1$, $\lambda = 2\times 10^{-10}$, $g = 0.0288$, $\mu = 0.078$, $k = 10^{-3} ~l_p^{-1}$. The cosmic time $t$ is in Planck time units.}
\label{fig:R-1}
\end{figure}

\begin{figure}[!h]
\centering
\subfigure{
\begin{minipage}[b]{0.45\textwidth}
\includegraphics[width=\textwidth,keepaspectratio=true]{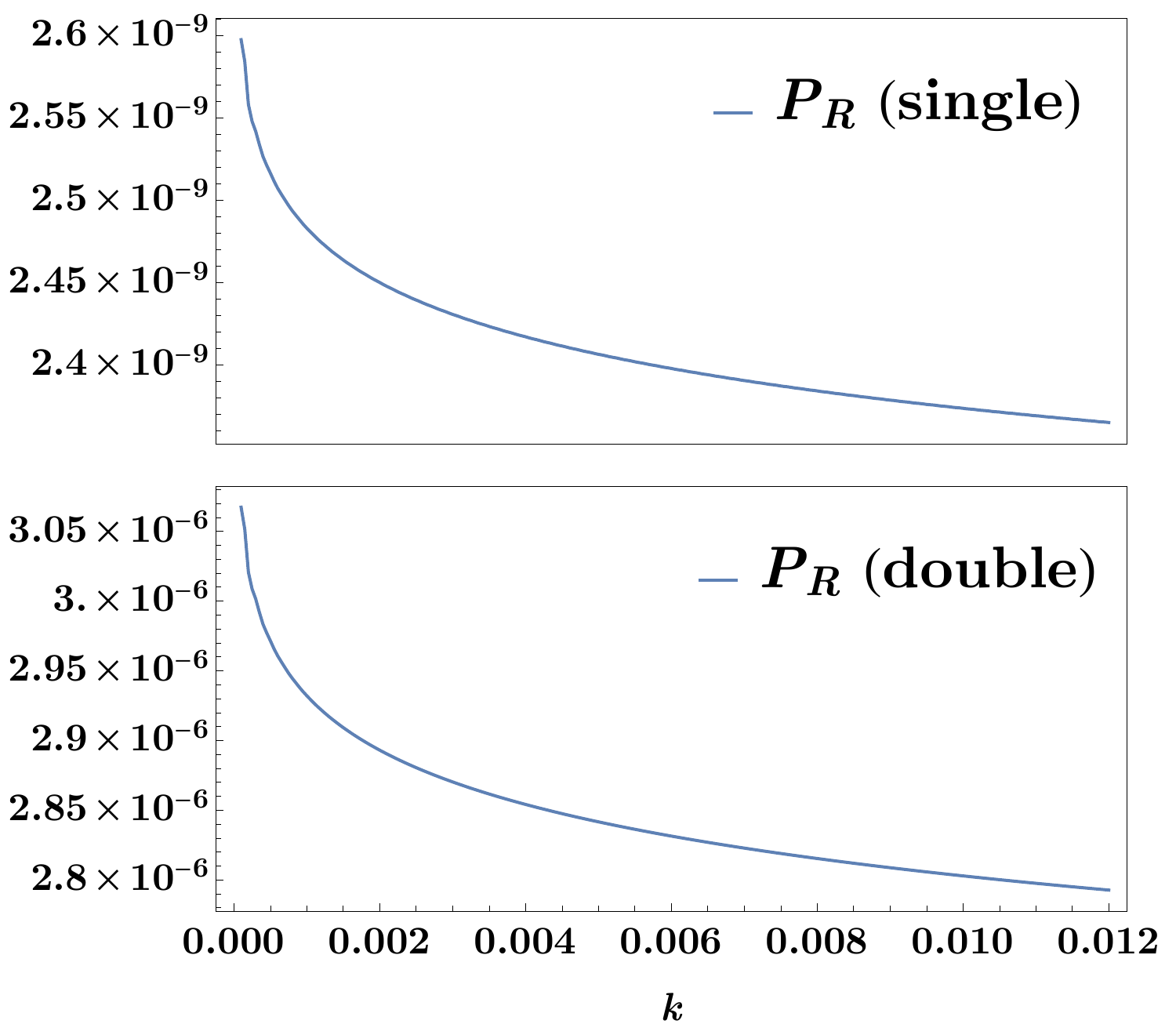}
\end{minipage}
}
\caption{Power spectra of single field curvature perturbation (upper) and double field one (lower) with model parameters $p=2/3$, $\lambda = 3.2 \times 10^{-10}$, $g = 0.057$, $\mu = 0.058$. The comoving wavenumber $k$ is in Planck units.}
\label{fig:power-2/3}
\end{figure}

Figure~\ref{fig:R-1} shows that the double fields curvature perturbation is not frozen after exiting the Hubble horizon, unlike the single field curvature perturbation. The peaks correspond to the particle production. The curvature perturbation has an overall increase on average.

Figure~\ref{fig:power-2/3} is the case of $p = 2/3$ showing that the second field makes such a significant contribution to lift the power spectrum that it cannot simply be neglected. At the late time of preheating, $\phi^2$ dominates so the power spectrum has similar results as the $p=2$ model. The matter field perturbation contributes significantly to the overall curvature perturbation. In general, the power spectrum of the double field curvature perturbation for $p=1$, $2/3$, $1/2$ and $1/3$ all have an amplification of several magnitudes compared to that of single field inflationary models. In this regard, our present study provides a concrete realization of the curvaton mechanism via the preheating process for a generic inflation model with an arbitrary power-law potential.

Our results demonstrate the significance of possible matter field contributions to the curvature perturbation. The traditional $(n_s-r)$ plot used to constrain inflationary models do not take into account the preheating process (largely due to uncertainties in the details of preheating mechanics). However, our analysis demonstrates that preheating may strongly affect the curvature perturbation, radically altering the predictions of single field inflation models.

\section{Conclusion and Discussions}\label{Sec:conclusions}

In the present article, we developed a code based on the C++ computer language and MPI parallelization, aiming at the numerical simulation of the preheating process after inflation. Users can easily change the potential for the inflaton $\phi$ and matter $\chi$ fields. Here we have mainly focused on the non-perturbative particle producing process and its influences on the curvature perturbation.

We showed that the case of $p=2$ corresponds to the $\phi^2$ model studied in the literature, as a viability test for the code. We then extended the analysis to the small $p$ case where the extra term $\mu$ in our potential smooths the singularity. We find these $\phi^p$ models ultimately transition to the familiar $\phi^2$ model so that after choosing the proper parameters, particle production remains efficient. Because of this, our method can be used to study preheating in axion-monodromy inflationary models, by eliminating the singularity at the minimum of the potential, providing a smooth transition from the axion-monodromy like potential inflation to an ordinary $\phi^2$ potential, making the preheating process as efficient as the quadratic case.

Taking into account the effect of the matter field fluctuations, the curvature perturbation can keep evolving at super-Hubble scales, which conflicts with the single field curvature prediction. The extra field can introduce an entropy perturbation that acts as a source of the curvature perturbation, amplifying the power spectrum. The amplification can be large, leading to additional constraints on the parameter space. This can be regarded as a particular realization of the curvaton mechanism, in which the primordial power spectrum is dominated by the fluctuations of the curvaton field instead of those of the inflaton field. However, unlike the traditional curvaton paradigm which requires a process of curvaton decay, in our model, the preheating phase itself serves this purpose. Our study demonstrates that the ordinary predictions for the $(n_s-r)$ plot which are based on the single field inflationary phase may fail to capture crucial information due to the unavoidable preheating phase. We conclude that the preheating phase needs to ultimately be taken into account when we confront various inflationary models with cosmological observations in high accuracy. This fact calls for significant improvement in our fundamental understanding of the production of particles in the early universe. To gather more information connecting inflationary cosmology with cosmological observations, the detailed investigation on primordial gravitational waves during preheating process must be included, which has been extensively studied in the literature \citep{Khlebnikov:1997di, Easther:2006gt, Easther:2006vd, Easther:2007vj, GarciaBellido:2007dg, GarciaBellido:2007af, Dufaux:2007pt, Price:2008hq, Martin:2013tda}. We would like to leave the extension of this analysis based on our numerical code for a follow-up study.

\acknowledgments
\section*{Acknowledgments}
We are grateful to J. Chen, J. Delabrouille, Z. Huang, J. Liu, X. Tong, L. Xue, D.-G. Wang, Z. Wang and S.-Y. Zhou for valuable discussions.
YFC and JJ are supported in part by the NSFC (Nos. 11653002, 11722327, 11421303, J1310021), by National Youth Thousand Talents Program of China, by the CAST Young Elite Scientists Sponsorship Program (2016QNRC001), and by the Fundamental Research Funds for the Central Universities.
QYL is supported in part by the Department of Physics \& Astronomy at the University of Pennsylvania.
YZ is supported in part by the NSFC (Nos. 11675165, 11633001), the CAS Strategic Priority Research Program ``The Emergence of Cosmological Structures'' (No. XDB09000000) and SRFDP.
All numerical computations are operated on the computer clusters LINDA \& JUDY in the particle cosmology group at USTC.

\appendix
%\section{Initial conditions}
\section{Initial condition for the matter field $\chi$}
\label{app:chi}

We introduce the initial condition for the matter field $\chi$ at the end of the inflationary period, when $\epsilon=1$. We choose the Bunch-Davies vacuum to be the initial condition for cosmological perturbations.

We rewrite the equation of motion of the matter field by eliminating the first order derivative,
\begin{equation}
 \ddot{X}_k+\omega_k^2X_k=0
\end{equation}
where $X_k(t)=a^{\frac{3}{2}}(t)\chi_k(t)$ the $\omega_k$ is
\begin{align}
 \omega_k^2&=\frac{k^2}{a^2}+g^2\phi^2-\frac{9}{4}H^2-\frac{3}{2}\dot{H}\\
 &=\frac{k^2}{a^2}+g^2\phi^2+\frac{3}{8M_p^2}\dot{\phi}^2-\frac{3}{4}V(\phi).
\end{align}
so the Bunch-Davies vacuum in the rescaled coordinate is
\begin{equation}
 X_k=\frac{1}{\sqrt{2\omega_k}} e^{-i\omega_k t}=\frac{1}{\sqrt{2\omega_k}} (\cos\omega_k t-i\sin\omega_k t)
\end{equation}
the derivative is
\begin{align}
 \dot{X}_k&=\frac{-i\omega_k}{\sqrt{2\omega_k}} e^{-i\omega_k t}=(\frac{-i\omega_k}{\sqrt{2\omega_k}})(\cos\omega_k t-i\sin\omega_k t)\nonumber\\
 &=-\omega_k\sin\omega_k t-i\omega_k\cos\omega_k t,
\end{align}
so in the physical coordinate (real part) the corresponding vacuum is
\begin{equation}
 \chi_k = a^{-3/2}\frac{1}{\sqrt{2\omega_k}} e^{-i\omega_k t} = a^{-3/2}\frac{1}{\sqrt{2\omega_k}} \cos\omega_k t,
\end{equation}
the derivative is
\begin{align}
 \dot{\chi}_k & = -\frac{3}{2} a^{-5/2} \dot{a} \frac{1}{\sqrt{2\omega_k}} e^{-i\omega_k t} + a^{-3/2}\frac{-i\omega_k}{\sqrt{2\omega_k}} e^{-i\omega_k t}\nonumber\\
 & = -\frac{3}{2} a^{-3/2} H \frac{1}{\sqrt{2\omega_k}} \cos\omega_k t - a^{-3/2} \frac{\omega_k}{\sqrt{2\omega_k}} \sin\omega_k t\nonumber\\
 & = -a^{-3/2} \frac{1}{\sqrt{2\omega_k}} (\frac{3}{2} H \cos\omega_k t + \omega_k \sin\omega_k t)
\end{align}

\section{Initial conditions for cosmological perturbations $Q_{\phi}$ and $Q_{\chi}$}
\label{app:Q}

Similar to the former subsection, we rescale the equation of cosmological perturbations in comoving time
\begin{align}
 &\ddot{Q}_{\phi}+3H\dot{Q}_{\phi}+\frac{k^2}{a^2}Q_{\phi}+M_{\phi\phi}Q_\phi+M_{\phi\chi}Q_\chi=0\nonumber\\
 \Rightarrow\qquad &Q^{\prime\prime}_\phi+2\mathcal{H}Q^{\prime}_{\phi}+k^2Q_{\phi}+a^2M_{\phi\phi}Q_\phi+a^2M_{\phi\chi}Q_\chi=0,
\end{align}
where the prime denotes the derivative with respect to the comoving time, $\mathcal{H}$ is the Hubble parameter in comoving time, $M_{\phi\phi}$ is given by $V_{,\phi\phi}-\frac{8\pi G}{a^3}\frac{d}{dt}\bigg(\frac{a^3}{H}\dot{\phi}\dot{\phi}\bigg)$, $M_{\phi\chi}$ is $V_{,\phi\chi}-\frac{8\pi G}{a^3}\frac{d}{dt}\bigg(\frac{a^3}{H}\dot{\phi}\dot{\chi}\bigg)$. Initially the matter field $\chi$ is absent, and hence $M_{\phi\chi}=0$.

We introduce the canonical variable $\nu_\phi=a Q_\phi$, and then derive the equation without first order derivative
\begin{equation}
 \nu^{\prime\prime}_\phi-\mathcal{H}^{\prime}\nu_\phi-\mathcal{H}^2\nu_\phi+k^2\nu_\phi+M_{\phi\phi}\nu_\phi=0,
\end{equation}
and thus, the Bunch-Davies vacuum for the perturbation of the Mukhannov-Sasaki variable takes $\nu_\phi = a Q_\phi \rightarrow \frac{e^{-i\omega_k\tau}}{\sqrt{2\omega_k}} $, where $\omega_k = k^2 -\mathcal{H}^{\prime} -\mathcal{H}^2+M_{\phi\phi}$, and the derivative is
\begin{align}
 \nu_\phi^\prime =& (aQ_\phi)^\prime \rightarrow -i\sqrt{\frac{\omega_k}{2}} e^{-i\omega_k\tau} \nonumber\\
 =& a^\prime Q_\phi+aQ_\phi^\prime = a^2HQ_\phi+a^2\dot{Q}_\phi \rightarrow -i \sqrt{\frac{\omega_k}{2}} e^{-i\omega_k\tau} ~.
\end{align}
Accordingly, in physical scale, the initial condition for the perturbation takes
\begin{equation}
 Q_\phi^{\text{ini}}\rightarrow\frac{1}{a\sqrt{2\omega_k}}e^{-i\omega_k\tau},
\end{equation}
and the associated derivative term is expressed as
\begin{align}
 \dot{Q}_\phi^{\text{ini}} &\rightarrow -\frac{i}{a^2} \sqrt{\frac{\omega_k}{2}} e^{-i\omega_k\tau} -H_{\text{ini}} Q_\phi^{\text{ini}} \nonumber\\
 &= -\frac{i}{a^2} \sqrt{\frac{\omega_k}{2}} e^{-i\omega_k\tau} -H_{\text{ini}} \frac{1}{a\sqrt{2\omega_k}} e^{-i\omega_k\tau} \nonumber\\
 &= \frac{1}{a\sqrt{2\omega_k}} e^{-i\omega_k\tau} \Big( \frac{-i\omega_k}{a}-H_{\text{ini}} \Big) ~.
\end{align}
Both real and imaginary part should be taken into consideration.

\end{document}